\documentstyle[12pt,bezier]{article}
\def\sq{\mathchoice{\square{6pt}}{\square{5pt}}{\square{4pt}}{\square{3pt}}}
\def\square#1{\mathop{\mkern0.5\thinmuskip\vbox{\hrule\hbox{\vrule\hskip#1 \vrule height#1 width0pt\vrule}\hrule}\mkern0.5\thinmuskip}}
\def\ni{\noindent}
\def\be{\begin{equation}}
\def\ee{\end{equation}}
\def\bea{\begin{eqnarray}}
\def\eea{\end{eqnarray}}

\begin{document}
\bibliographystyle{unsrt}

\begin{titlepage}

\null
\vskip-3truecm
\hspace*{5truecm}{\hrulefill}
\par
\vskip-3truemm
\par
\hspace*{5truecm}{\hrulefill}
\par
\vskip5mm
\par
\hspace*{5truecm}{{\large\sf University of Greifswald (April, 
1995)}}
\vskip4mm
\par
\hspace*{5truecm}{\hrulefill}
\par
\vskip-3truemm
\par
\hspace*{5truecm}{\hrulefill}
\par
\bigskip
\hspace*{5truecm}hep-th/yymmnn
\bigskip
\par
\hspace*{5truecm}Submitted to: {\sf Nucl.Phys. B}

\vskip3cm

\centerline{\Large\bf A model of stable chromomagnetic vacuum}
\medskip
\centerline{\Large\bf in higher-dimensional
Yang-Mills theory}

\bigskip
\centerline{\Large\sf Ivan G. Avramidi
\footnote{On leave of absence from Research Institute
for Physics, Rostov State 
University,  Stachki 194, 344104 Rostov-on-Don,
Russia.}
\footnote{\sc E-mail: 
avramidi@math-inf.uni-greifswald.d400.de}}
\medskip
\centerline{\it Department of Mathematics, University of Greifswald}
\centerline{\it Jahnstr. 15a, 17489 Greifswald, Germany}

\medskip
\centerline{\today}
\medskip

\begin{abstract}

We study the effective action in Euclidean Yang-Mills theory with 
a compact simple gauge group in one-loop approximation assuming a covariantly constant gauge field strength as a background. 
For groups of higher rank and spacetimes of higher dimensions such field configurations have many independent color components taking values in Cartan subalgebra and many ``magnetic fields'' in each color component.
In our previous
investigation it was shown that such background is stable in 
dimensions
higher than four provided the amplitudes of ``magnetic fields'' do not differ much from each other. 
In present paper we
calculate exactly the relevant 
zeta-functions in the case of equal amplitudes of ``magnetic fields''. For the case of two ``magnetic fields'' with equal amplitudes the behavior of the effective action is studied in detail.
It is shown that in dimensions $d=4,5,6,7$ $({\rm mod}\, 8)$, the perturbative vacuum is metastable, i.e., it is 
stable in 
perturbation theory but  the effective action is not bounded from below, whereas in 
dimensions $d=9,10,11$ $({\rm mod}\, 8)$ the perturbative vacuum is absolutely stable. In 
dimensions $d=8$ $({\rm mod}\, 8)$ the perturbative vacuum is stable for small values of 
coupling constant but becomes unstable for large coupling constant leading to the 
formation of a 
non-perturbative stable vacuum with nonvanishing ``magnetic fields''. 
The 
critical value of the coupling constant and the amplitudes of the vacuum ``magnetic fields''
are evaluated exactly. 
 
\end{abstract}



\end{titlepage}


\section{Introduction}

The low-energy behavior of the non--Abelian gauge field theories is an open problem of 
quantum field theory. At low energies the effective interactions 
become strong and break down 
the usual perturbation theory, which works very well in the high-energy limit. 
This means that the vacuum at low energies has a far more complicated 
structure than the perturbative 
high--energy one.

To analyze the infra-red behavior of a $SU(2)$ Yang-Mills theory 
Savvidy \cite{Savvidy} put forward an explicit ansatz for the vacuum gauge fields in 
form of a covariantly constant gauge field strength.
Such a field configuration is 
necessarily Abelian, i.e., it takes values
in the Cartan subalgebra, and, hence, since 
${\rm rank }\, SU(2)=1$, has the only nonvanishing color component. On the other hand 
this component has the form of a constant magnetic field. Therefore such a 
configuration is usually called {\it constant chromomagnetic field}. 
Savvidy showed that due 
to the quantum fluctuations of the gauge fields the energy of such a field 
configuration is below the perturbative vacuum level, which leads to infra--red instability 
of the perturbative vacuum under creation of the constant chromomagnetic field. Further 
investigations \cite{Pagels,Nielsen78,Nielsen81,Olesen,Abbot,Consoli,Preparata}
showed that the constant chromomagnetic vacuum itself is unstable too, 
meaning that the real physical nonperturbative vacuum has even more complicated
structure.

In our recent papers \cite{Avr-jmp95a,Avr-leipz} we extended the Savvidy's investigation
by considering more complicated gauge groups and spacetimes of dimensions higher than four. 
We showed that there exist more general nontrivial field configurations that turn out to be stable.

In present paper we are going to consider 
some explicit example of such 
background field configurations when the problem can be solved {\it exactly} to the end.


\section{Infra-red regularization of the low-energy effective action}

We consider the pure Yang--Mills model with an arbitrary simple 
compact gauge group in an {\it Euclidean} $d$-dimensional flat spacetime.
The classical action of the model is
\be
S=\int dx\,{1\over 2g^2}{\rm tr}_{Ad}\,{\cal F}_{\mu\nu}{\cal F}^{\mu\nu},
\ee
where $g$ is the coupling constant, ${\cal F}_{\mu\nu}$ is the gauge field strength taking values in the Lie algebra of the gauge group and ${\rm tr}_{Ad}$ means the trace in adjoint representation.

In the case of covariantly constant background fields,

\be
\nabla_\mu{\cal F}_{\alpha\beta}=0,
\ee
the Euclidean effective 
action does not depend on the gauge \cite{Avr-jmp95a} and is usually given in one-loop approximation 
by (see e.g. \cite{DeWitt})

\be
\Gamma=S+\hbar\Gamma_{(1)}+O(\hbar^2),
\ee

\begin{equation}
\Gamma_{(1)}
={1\over 2}\log\,{\rm Det}\,{\Delta\over\mu^2}
-\log\,{\rm Det}\,{D\over\mu^2}.
\label{1}
\end{equation}

\noindent
Here `Det' means the functional determinant, $\mu$ is a dimensionful renormalization parameter and
$\Delta$ and $D$ are elliptic second-order differential operators of Laplace type

\begin{equation}
\Delta^\mu_{\ \nu}
=-\Box\delta^\mu_{\ \nu}
-2{\cal  F}^\mu_{\ \,\nu},
\label{2}
\end{equation}
\begin{equation}
D=-\Box,
\label{3}
\end{equation}

\noindent
where $\Box=\nabla_\mu\nabla^\mu$ is the covariant Laplacian, the covariant derivative $\nabla_\mu$ being taken in adjoint representation.

In practice, to study the infra--red behavior of the system one has to be a bit more careful.
Namely, to be able to carry out the intermediate calculations one should impose some 
infra--red regularization by introducing, say, a regularization mass parameter $M$
and take it off at the very end. 
The regularization parameter $M$ must be 
sufficiently large to ensure the infra--red convergence. Then one has to do the
analytical continuation to the neighborhood of the point $M=0$ and to take the limit 
$M\to 0$.
If there are no infra--red 
divergences then this procedure is harmless. In the case of non--trivial low-energy
behavior there appears an imaginary part 
of the effective action or some infra-red 
logarithmic singularities, so that in addition to the ultra-violet renormalization it could also be necessary to do the infra-red renormalization.

The infra--red regularized effective action $\Gamma(M)$ is not analytic, in general, at the vicinity of the point $M=0$.
Therefore, one should assume the infra--red regularization parameter 
to have an
infinitesimal negative imaginary part, i.e.,
$M^2\to M^2-i\varepsilon$. This enables one to choose the correct way of analytical 
continuation leading to the correct sign of the imaginary part of the effective action.
Thus, strictly speaking, the effective action should be defined by

\begin{equation}
\Gamma\stackrel{\rm def}{=}
\lim_{M\to 0-i\varepsilon}\Gamma(M),
\label{4}
\end{equation}

\noindent

In the one-loop approximation one can define the infra-red regularized effective action by

\begin{equation}
\Gamma_{(1)}(M)
={1\over 2}\log\,{\rm Det}\,{(\Delta+M^2)\over\mu^2}
-\log\,{\rm Det}\,{(D+M^2)\over\mu^2},
\label{5}
\end{equation}

\noindent
and suppose $M^2$ to be large enough, so that the operators $\Delta+M^2$ and $D+M^2$ 
are positive, i.e., do not have negative modes.

The functional determinants are also known to be ultra--violet divergent. To regularize 
the ultra--violet divergences one can take any usual regularization, e.g. the 
dimensional regularization or the zeta-function one etc. 
One should stress that the meaning of the
infra--red regularization is very different from the ultraviolet one. Although the ultraviolet regularization is purely formal method to get 
physically predictable results, the infra--red regularization parameter can take, in 
principle, a {\it direct physical meaning}, something like $\Lambda_{QCD}$. 
Anyway, the infra-red regularization should be taken off at the very end, {\it after} 
the ultraviolet regularization.


\section{Zeta--function and heat kernel}

Within the zeta-function regularization \cite{Dowker,Hawking,Elizalde} 
the effective action has the form

\begin{equation}
\Gamma_{(1)}(M)
=-{1\over 2}\zeta'_{\rm YM}(0, M),
\label{6}
\end{equation}

\ni
where

\begin{equation}
\zeta_{\rm YM}(p, M)\stackrel{\rm def}{=}
\zeta_{\Delta+M^2}(p) - 2\zeta_{D+M^2}(p)
\label{7}
\end{equation}

\ni
is the infra-red regularized total Yang-Mills zeta-function.

The zeta-function of an elliptic differential operator $L$ can be defined in terms of the associated heat kernel as follows 
\cite{Dowker,Hawking,Elizalde}

\begin{equation}
\zeta_L(p)=\mu^{2p}{\rm Tr}L^{-p}
=\int d\,x\,{\mu^{2p}\over \Gamma (p)}
\int\limits_0^\infty d\,t\,t^{p-1}{\rm tr}\,U_L(t),
\label{8}
\end{equation}

\ni
where `Tr' means the functional trace, `tr' is the usual  matrix trace and

\begin{equation}
U_L(t)=\exp(-t L)\delta(x,x')\Big\vert_{x=x'}
\label{10}
\end{equation}

\ni
is the heat kernel diagonal.
Therefore, the Yang--Mills zeta--function is expressed in terms of the heat kernel diagonals of the operators $\Delta$ and $D$

\begin{equation}
\zeta_{\rm YM}(p,M)
=\int d\,x\,{\mu^{2p}\over \Gamma (p)}
\int\limits_0^\infty d\,t\,t^{p-1}e^{-t M^2}U_{\rm YM}(t),
\label{11}
\end{equation}

\ni
where

\begin{equation}
U_{\rm YM}(t)={\rm tr}_{Ad}\,\{{\rm tr}_{T}\,U_\Delta(t)-2U_D(t)\}
\label{12}
\end{equation}

\ni
and `${\rm tr}_{T}$' means the trace over the vector indices.

The heat kernel diagonals of Laplace type operators $L=-\sq+P$ on covariantly constant background are calculated in
\cite{Avr-jmp95a} using an algebraic approach developed in \cite{Avr-plb93}. 
This approach was developed further, also to the case of curved manifolds, in 
\cite{Avr-jmp95b,Avr-plb94,Avr-jmp96a}. 
(For review see \cite{Avr-win,Avr-qgr6,Avr-thess}).

By using the results of \cite{Avr-jmp95a} we have

\begin{equation}
U_{\rm YM}(t)=(4\pi t)^{-d/2}{\rm tr}_{Ad}\left\{
{\rm det}_{T}\left({t\hat {\cal  F}\over \sinh(t\hat {\cal  F})}\right)^{1/2}
\Biggl({\rm tr}_{T}\exp(2t\hat{\cal  F})-2\Biggr)\right\},
\label{13}
\end{equation}

\ni
where  $\hat{\cal F}=\{{\cal F}^\mu_{\ \,\nu}\}$ is a matrix with vector indices and
`${\rm det}_{T}$' means the determinant over the vector indices.

To calculate the trace over the group indices we observe first 
that covariantly constant 
Yang--Mills fields are necessarily Abelian,

\begin{equation}
[{\cal F}_{\mu\nu}, {\cal F}_{\alpha\beta}]=0,
\label{14}
\end{equation}

\ni
and, therefore, take their values in the Cartan subalgebra.
Thus the maximal number of 
independent fields is equal to the dimension of the Cartan subalgebra, i.e., to the 
rank of the group $r$.
Using the property (\ref{14}) one obtains \cite{Avr-jmp95a}

\begin{equation}
U_{\rm YM}(t)=(4\pi t)^{-d/2}
\Biggl\{r(d-2)
+2\sum_{\alpha>0}
{\rm det}_{T}\left({t \hat F\over \sin(t\hat  F)}\right)^{1/2}
\Biggr({\rm tr}_{T}\cos(2t \hat F)-2\Biggl)\Biggr\},
\label{15}
\end{equation}

\noindent
where 
\begin{equation}
\hat F=\{ F^{a\,\mu}_{\ \ \ \nu}\alpha_a\},
\label{16}
\end{equation}

\ni
$\alpha_a$ are the positive roots of the Lie algebra of the gauge group and the sum runs
over all positive roots. The number of positive roots is $p=(n-r)/2$, where $n$ is the dimension of the gauge group.

Further, by orthogonal transformations one can always put the antisymmetric $d\times d$ matrix 
$\hat F$ in the block-diagonal form
with $q$ two-dimensional antisymmetric blocks on the diagonal, all others 
entries being zero

\begin{equation}
\hat F=\left(
\begin{array}{cccccccccc}
0    	&H_1	&0	&0	&\cdots	&0	&0	&0 	&\cdots	&0	\\
-H_1 	&0 	&0 	&0 	&\cdots &0	&0 	&0	&\cdots	&0	\\
0	&0  	&0	&H_2	&\cdots	&0	&0 	&0 	&\cdots	&0	\\
0	&0  	&-H_2	&0	&\cdots	&0	&0 	&0 	&\cdots	&0	\\
\vdots	&\vdots	&\vdots	&\vdots &\ddots	&\vdots &\vdots &\vdots &\cdots &\vdots	\\
0	&0	&0  	&0	&\cdots &0	&H_q	&0 	&\cdots &0	\\
0	&0	&0  	&0	&\cdots &-H_q	&0	&0 	&\cdots &0	\\
0	&0	&0  	&0	&\cdots &0 	&0 	&0	&\cdots &0	\\
\vdots	&\vdots	&\vdots &\vdots	&\cdots &\vdots	&\vdots	&\vdots	&\ddots	&\vdots	\\
0	&0	&0  	&0	&\cdots &0 	&0 	&0 	&\cdots	&0
\end{array}
\right).
\label{17}
\end{equation}

\ni
Visually one can think of such field configuration as a number of `magnetic fields' in different directions and 
call the invariants $H_i$ the amplitudes of `magnetic fields'. They can be expressed in terms of the basic invariants of the matrix $\hat F$,

\begin{equation}
I_k = {1\over 2}(-1)^k{\rm tr}_{T} 
\hat F^{2k},
\label{20}
\end{equation}

\ni
by solving the equations

\begin{eqnarray}
\left\{
\begin{array}{lllllcl}
 H_1^2	&+	&\cdots	& +	&H_q^2		&=	&I_1\\
\vdots	&	&		&	&\vdots		&&\vdots\\
H_1^{2[d/2]}&+&\cdots	&+	&H_q^{2[d/2]}	&=	&I_{[d/2]}
\end{array}.
\right.
\label{19}
\end{eqnarray}

One should note that, although the matrix $\hat F$ depends linearly on the roots, its invariants $H_i$ depend on the roots in a very nontrivial way.

By making use of (\ref{17}) one can compute the trace and the
determinant over 
the vector
indices in eq. (\ref{15}) and obtain finally 
 
\begin{eqnarray}
\lefteqn{U_{\rm YM}(t)=(4\pi t)^{-d/2}\Biggl\{r(d-2)}
\nonumber\\
&&+ 2\sum_{\alpha>0;}
\prod_{1\le i\le q}\left({tH_i\over \sinh(tH_i)}\right)
\left(d-2 
+ 4\sum_{1\le j\le q}\sinh^2(tH_j)\right)\Biggr\}.
\label{21}
\end{eqnarray}

\ni
Hence the total zeta-function for the gauge fields, (\ref{11}), 
equals

\begin{eqnarray}
\lefteqn{\zeta_{YM}(p,M) =
\int dx (4\pi)^{-d/2}{\mu^{2p}\over \Gamma(p)}
\int\limits_0^\infty d t\,t^{p-d/2-1}e^{-t M^2}
\Biggl\{r(d-2)}
\qquad\ \
\nonumber\\
&&+2\sum_{{ \alpha}>0;}
\prod_{1\le i\le q}\left({tH_i\over 
\sinh(tH_i)}\right)
\left(d-2 + 4\sum_{1\le j\le 
q}\sinh^2(tH_j)\right)\Biggr\}.
\label{22}
\end{eqnarray}

For ${\rm Re}\, p>d/2$ the integral (\ref{22}) over $t$ converges at $t\to 0$. 
For sufficiently large $M$ it also converges at $t\to\infty$ for arbitrary values of amplitudes of `magnetic fields' $H_i$.
It is easy to see, however, that for $M=0$ there are some field configurations that break down the convergence of this integral at $t\to\infty$. The simplest example is the case of only one `magnetic field', i.e., $q=1$.

In \cite{Avr-jmp95a} it was found a 
criterion  of infra--red stability, i.e., the convergence of the integral (\ref{22}) over $t$ at $t\to\infty$ in the limit $M\to 0$. 
Roughly speaking, the integral (\ref{22}) converges 
when the amplitudes of `magnetic fields' $H_i$ do 
not differ much from each other.
More precisely, the vacuum is infra--red stable, i.e., the integral (\ref{22}) converges at $t\to\infty$ for $M=0$, when the background `magnetic fields' satisfy the {\it condition of
stability},

\begin{equation}
\max_{1\le i\le q} H_i< \sum_{1\le i\le q}^{\ \quad\prime} H_i
\qquad {\rm for\ any\ root\ } \alpha,
\label{25}
\end{equation}

\ni
where the prime at the sum means that the summation does not 
include the maximal invariant.

%

Obviously, the condition (\ref{25}) can be fulfilled only in the case when the number of `magnetic fields' is not less than two, $q\ge 2$, and, hence, only in dimensions not less than four, $d\ge 4$. The case $q=2$ in $d=4$ is the critical one
(see the detailed discussion below).
That is why we also study in present 
paper the higher--dimensional case ($d\ge 5$) in detail. 
Even in this case the vacuum can be 
unstable. Since the stability is supported by approximately \hbox{equal} amplitudes of `magnetic fields', we 
study in this paper the `most stable' configuration when all the `magnetic fields' have equal amplitudes:

\begin{equation}
H_i=H, \qquad (i=1,\dots, q).
\label{26}
\end{equation}

\ni
This assumes certain algebraic relations between the basic invariants $I_k$ of the matrix $\hat F$

\be
H=\left({{I_1\over q}}\right)^{1/2}=\left({I_2\over q}\right)^{1/4}
=\cdots=\left({I_{[d/2]}\over q}\right)^{1/(2[d/2])}.
\ee

\ni
In this case the Yang-Mills zeta-function takes the form

\begin{eqnarray}
\zeta_{YM}(p,M) &=&
\int dx (4\pi)^{-d/2}{\mu^{2p}\over \Gamma(p)}
\int\limits_0^\infty d t\,t^{p-d/2-1}e^{-t M^2}
\Biggl\{r(d-2)\nonumber\\
&&+2\sum_{{ \alpha}>0}
\left({tH\over \sinh(tH)}\right)^q
\left(d-2 + 4q\,\sinh^2(tH)\right)\Biggr\}.
\label{27}
\end{eqnarray}

For sufficiently large $M$ the integral (\ref{27}) converges in the region ${\rm Re}\, 
p>d/2$ and defines an analytic function. Therefore, changing the integration variable 
$t\to t/H$ we get

\begin{eqnarray}
\zeta_{YM}(p,M)&=&\int dx (4\pi)^{-d/2}\mu^{2p}
{\Gamma(p-d/2)\over\Gamma(p)}
\Biggl\{r(d-2)M^{d-2p}
\nonumber\\&&
+2\sum_{{ \alpha}>0}H^{d/2-p}J_{d/2-p,\,q}\left({M^2\over H}\right)\Biggr\},
\label{28}
\end{eqnarray}

\ni
where

\be
J_{s,\,q}(z)=(d-2)b_{s,\,q}(z)+4q\,s(s-1)b_{s-2,\,q-2}(z),
\label{29}
\ee
\be
b_{s,\,q}(z)={1\over\Gamma(-s)}\int\limits_0^\infty d t\,t^{-s-1+q}{e^{-tz}\over \sinh^q t}.
\label{29d}
\ee

Thus the regularized zeta-function is expressed in terms of the functions $b_{s,\,q}(z)$.
These functions are studied in detail in the Appendix.
It is well known that the coefficient at $(-\log\,\mu)$ in the effective action is determined by the zeta-function at zero.
From eq. (\ref{28}) we immediately get

\begin{equation}
\zeta_{\rm YM}(0,M)=0 \qquad {\rm for \ odd\ } d,
\label{32}
\end{equation}

\ni
and

\begin{equation}
\zeta_{\rm YM}(0,M)=
\int dx (4\pi)^{-d/2}
{(-1)^{d/2}\over\Gamma(d/2+1)}
\Biggl\{r(d-2)M^{d}
+2\sum_{{\alpha}>0}J_{d/2,\,q}\left({M^2\over H}\right)H^{d/2}
\Biggr\}
\label{34}
\end{equation}
for even $d$.

Let us take the limit $M\to 0-i\varepsilon$ here. The contribution of the `free' fields is 
proportional to $M^d$ and vanishes in this limit. 
Defining
\be
J_{s,\,q}\stackrel{\rm def}{=}\lim_{z\to 0-i\varepsilon}J_{s,\,q}(z)
\label{33aa}
\ee
and using the formulas of the Appendix we find that $J_{s,\,q}$ for odd positive $s=2k+1$ vanishes,
\be
J_{2k+1,\,q}=0, \qquad (k=0,1,2,\dots).
\label{34aa}
\ee
whereas for even positive $s=2k$
\be
J_{2k,\,q}=(d-2)a_{2k,\,q}+8q\,k(2k-1)a_{2k-2,\,q-2}
\label{34aaa}
\ee
with (see Appendix)
\be
a_{2k,\,q}=\left.\left({\partial\over\partial t}\right)^{2k}\left({t\over\sinh t}\right)^q\right|_{t=0}.
\ee

Therefore, we have from (\ref{34})

\be
\zeta_{\rm YM}(0,0)
=0 \qquad{\rm for\ odd}\ d\ge 3\ {\rm and\ odd}\ d/2\ge 1,
\label{36}
\ee
and
\be
\zeta_{\rm YM}(0,0)
=\int dx \sum_{{\alpha}>0}(4\pi)^{-d/2}
{2\over\Gamma(d/2+1)}J_{d/2,\,q}
H^{d/2}
\quad{\rm for\ even}\ d/2\ge 2.
\ee


\section{One-loop effective action}

Using the zeta-function (\ref{28}) we obtain from (\ref{6}) then the effective action:

\begin{equation}
\Gamma_{(1)}(M)=\int dx {1\over 2}(4\pi)^{-d/2}
{\pi(-1)^{(d-1)/2}\over\Gamma(d/2+1)}
\Biggl\{r(d-2)M^{d}
+2\sum_{{ \alpha}>0}J_{d/2,\,q}\left({M^2\over H}\right) H^{d/2}\Biggr\}
\label{33}
\end{equation}

\centerline{for odd $d$}

\ni
and

\begin{eqnarray}
\lefteqn{
\Gamma_{(1)}(M)=\int dx {1\over 2}(4\pi)^{-d/2}
{(-1)^{d/2}\over\Gamma(d/2+1)}
}
\nonumber\\
&&
\times\Biggl\{r(d-2)M^{d}
\Biggl(\log\,{M^2\over\mu^2}-\Psi(d/2+1)-{\bf C}\Biggr)
\nonumber\\
&&
+2\sum_{{ \alpha}>0}H^{d/2}\left[J'_{d/2,\,q}\left({M^2\over H}\right)
+J_{d/2,\,q}\left({M^2\over H}\right)\left(\log\,{H\over\mu^2}-\Psi(d/2+1)-{\bf C}\right)
\right]\Biggr\}
\nonumber\\
&&
\label{35}
\end{eqnarray}
\centerline{for even $d$,}

\ni
where 
\be
J'_{d/2,\,q}(z)= \left.{\partial J_{s,\,q}(z)\over \partial s}\right|_{s=d/2},
\ee
$\Psi(z)\equiv d \log \Gamma(z)/d\,z$ and ${\bf C}=-\Psi(1)$ is the Euler constant.

Very similar formulas for the zeta-function and the effective action were obtained in general case in 
\cite{Avr-npb91}
in a bit different context (see especially the eqs. (2.25), (2.28) and (2.29) in Sect. 2 therein).

Thus the effective action is determined by two quantities --- $J_{d/2,\,q}(M^2/H)$ and $J'_{d/2,\,q}(M^2/H)$, i.e., by the values of the function $J_{s,\,q}(z)$ at the positive integer and half-integer  points, $s=k/2$, and its derivative at integer points, $s=k$. Using the eq. (\ref{29}) and the properties of the function $b_{s,\,q}(z)$ studied in the Appendix, it is not difficult to obtain all these coefficients.

By taking off the infra-red regularization and using the eq. (\ref{34aa})
we obtain finally

\begin{equation}
\Gamma_{(1)}=\int dx \sum_{{ \alpha}>0}
(4\pi)^{-d/2}
{\pi(-1)^{(d-1)/2}\over\Gamma(d/2+1)}J_{d/2,\,q}
H^{d/2}
\ \ {\rm for\ odd}\ d\ge 3,
\label{37}
\end{equation}

\be
\Gamma_{(1)}=\int dx\sum_{{ \alpha}>0}
 (4\pi)^{-d/2}
{-1\over\Gamma(d/2+1)}J'_{d/2,\,q}H^{d/2}
\qquad {\rm for\ odd}\ d/2\ge 1,
\label{39}
\ee

and

\begin{eqnarray}
\Gamma_{(1)}&=&\int dx\sum_{{ \alpha}>0}
 (4\pi)^{-d/2}
{1\over\Gamma(d/2+1)}
\nonumber\\
&&
\times H^{d/2}\left\{J'_{d/2,\,q}
+J_{d/2,\,q}\left[\log\,{H\over\mu^2}-\Psi(d/2+1)-{\bf C}\right]\right\}
\label{39b}
\end{eqnarray}
\centerline{for even $d/2\ge 2$.}

Thus one has to calculate the values of the function $J_{s,\,q}$  and its derivative $J'_{s,\,q}$ at half-integer points and integer points. 
We distinguish three essentially different cases: i) $q=1$, ii) $q=2$, and iii) $q\ge 3$.

\subsection{One `magnetic' field}

The case $q=1$ is realizable in any dimensions $d\ge 2$. Therefore, we need to study the function $J_{s,\,1}$ in the region ${\rm Re}\,s\ge 2$. It should be noted, however, that  this is the only possible case in  four-dimensional spacetime of {\it Lorentzian} signature \cite{Avr-jmp95a}. It is this case that was studied by Savvidy \cite{Savvidy}.

Using the formulas of the Appendix we obtain from eqs. (\ref{29}) and (\ref{33aa}) in this case

\begin{equation}
J_{s,\,1}=4(d-2)(2^{s-1}-1)(2\pi)^{-s}\cos\left({\pi \over 2}s\right)\Gamma(s+1)\zeta(s)
+2s\left(e^{-i\pi s}+1\right).
\label{52a}
\end{equation}

\ni
Therefrom we obtain for odd dimensions, i.e., for half-integer $s=d/2$,

\bea
J_{d/2,\,1}&=&(-1)^{[(d+1)/4]}2(d-2)(2^{d/2-1}-1)\sqrt 2\,(2\pi)^{-d/2}\Gamma(d/2+1)\zeta(d/2)
\nonumber\\
&&
+d-i (-1)^{(d-1)/2}d,
\label{47a}
\eea
$$
{\rm for\ odd}\ d\ge 3, \ {\rm i.e., }\ d=3,5,7,\dots.
$$

\ni
For even dimensions, i.e., integer $s=d/2$, we find from (\ref{52a}) in accordance with eqs. (\ref{34aa}) and (\ref{34aaa}) 
\be
J_{d/2,\,1}=0 \qquad {\rm for\ odd}\ d/2\ge 1, {\rm i.e., }\ d=2,6,10, \dots
\ee
and, by using \cite{Erdelyi}
\be
\zeta(2k)=(-1)^{k+1}(2\pi)^{2k}{B_{2k}\over 2(2k)!},
\ee
where $B_k$ are the Bernulli numbers,

\be
J_{d/2,\,1}=-2(d-2)(2^{d/2-1}-1){B_{d/2}}+2d, 
\label{48}
\ee
$$ 
{\rm for \ even}\ d/2\ge 2, \ {\rm i.e.,}\ d=4,8,12,\dots.
$$

\ni
The derivative  $J'_{d/2,\,1}$ for even dimensions is
 
\be
J'_{d/2,\,1}
=-(-1)^{(d-2)/4}(d-2)(2^{d/2-1}-1)\Gamma(d/2+1){\zeta(d/2)\over (2\pi)^{d/2-1}}
+i\pi d
\label{49d}
\ee
$$
{\rm for \ odd}\ d/2\ge 1,\ {\rm i.e.,}\ d=2,6,10,\dots
$$

\begin{eqnarray}
J'_{d/2,\,1}
&=&-2(d-2)(2^{d/2-1}-1)B_{d/2}\Biggl[{\zeta'(d/2)\over \zeta(d/2)}
+\Psi(d/2+1) 
\nonumber\\
&&
+{2^{d/2-1}\over 2^{d/2-1}-1}\log 2-\log(2\pi)\Biggr]
+4-i\pi d
\nonumber\\
&&
\label{49}
\end{eqnarray}
$$
{\rm for \ even}\ d/2\ge 2, \ (d=4,8,12,\dots).
$$

Substituting (\ref{47a}), (\ref{49d})   and (\ref{49}) 
in (\ref{37})--(\ref{39b}) we find that 
the effective action gets a negative  
imaginary part

\begin{equation}
{\rm Im}\, \Gamma_{(1)}
=-\int dx \sum_{{ \alpha}>0}
(4\pi)^{-d/2}{2\pi\over \Gamma(d/2)}H^{d/2},
\qquad {\rm for }\ q=1,
\label{52}
\end{equation}

\ni
which indicates on the instability of the chromomagnetic vacuum with one `magnetic field', i.e., $q=1$.
It is remarkable that this form is valid for any dimension.


\subsection{Two `magnetic' fields}

Now let us consider the most interesting case of two `magnetic' fields, i.e., $q=2$, that is possible in ($d\ge 4$). 
Again we have to study the region ${\rm Re}\,s\ge d/2\ge 2$.
This case is {\it critical}
because for $q=1$ there is a strong infra--red instability, whilst for $q\ge 3$ the 
model is infra-red stable. The limit $z\to 0-i\varepsilon$ of the function $J_{s,\,2}(z)$ is not regular. From eqs. (\ref{29}) and  (\ref{33aa}) and the Appendix we have now

\bea
\left.J_{s,\,2}(z)\right|_{z\to 0-i\varepsilon}
&=&2^{s+1}(d-2)(s-1)(2\pi)^{-s}
\cos\left({\pi \over 2}s\right)\Gamma(s+1)\zeta(s)
\nonumber\\
&&
+ 8s(s-1){z}^{s-2}.
\label{51}
\eea

Note that the limit $z\to 0-i\varepsilon$ is well defined only in the region ${\rm Re}\,s> 2$, i.e., for $d\ge 5$.
In this region we have

\be
J_{s,\,2}=2^{s+1}(d-2)(s-1)(2\pi)^{-s}\cos\left({\pi \over 2}s\right)\Gamma(s+1)\zeta(s).
\label{62c}
\ee

\ni
Therefore, 

\be
J_{d/2,\,2}=(-1)^{[(d+1)/4]}2^{(d-1)/2} (d-2)^2(2\pi)^{-d/2}\Gamma(d/2+1)\zeta(d/2), 
\ee
$$
{\rm for\ odd}\ d\ge 5, \ (d=5,7,9\dots)
$$
and
(in accordance with (\ref{34aa}) and (\ref{34aaa}) )
\begin{equation}
J_{d/2,\,2}=0 \qquad {\rm for\ odd}\ d/2\ge 3,\ (d=6,10,14,\dots)
\ee

\be
J_{d/2,\,2}=-2^{d/2-1}(d-2)^2B_{d/2} \qquad {\rm for\ even}\ d/2\ge 4,\ (d=8,12,16,\dots).
\label{57c}
\end{equation}

\ni
The derivative $J'_{d/2,\,2}$ for even dimensions is given by

\be
J'_{d/2,\,2}=(-1)^{[(d+2)/4]}2^{d/2-2}(d-2)^2\Gamma(d/2+1){\zeta(d/2)\over (2\pi)^{d/2-1}}
\ee
$$
{\rm for\ odd}\ d/2\ge 3,\ (d=6,10,14,\dots)
$$
and
\be
J'_{d/2,\,2}=-2^{d/2-1}(d-2)^2B_{d/2}
\Biggl[{\zeta'(d/2)\over \zeta{d/2}}
+\Psi(d/2+1)+\log 2
-\log(2\pi)+{2\over d-2}\Biggr]
\label{49e}
\ee
$$
{\rm for\ even}\ d/2\ge 4,\ (d=8,12,16,\dots).
$$

The case $d=4$ is not regular one because the point $s=2$ is singular in the limit $z\to 0$. 
The value $J_{2,\,2}$ is still finite

\be
J_{2,\,2}=-{4\over 3}+16={44\over 3}, \qquad {\rm for}\ d=4,
\ee
where the second contribution, $+16$, comes from the $z$-\-de\-pen\-dent term in (\ref{51}). Note that this cannot be obtained from (\ref{57c})
by putting $d=4$, but is in fully accordance with eq. (\ref{34aaa}) for $q=2$ and $d=4$.

However, the derivative $J'_{d/2,\,2}(z)$ for $d=4$ diverges in the limit $z\to 0$

\be
\left.J'_{2,\,2}(z)\right|_{z\to 0}
=16\log{z}-{4\over 3}\left({\zeta'(2)\over \zeta(2)}-{\bf C}-\log (2\pi)+\log 2+1\right)
+22+o(z).
\ee
Therefore, from eq. (\ref{35}) we find that the effective action  $\Gamma(M)$ has a logarithmic infra--red divergency in the limit $M\to 0$
\be
\Gamma_{(1)}(M)\Big|_{M\to 0}=-\int dx\sum\limits_{\alpha}(4\pi)^{-2}8H^2\log{H\over M^2}+O(1),
\qquad (q=2,\ d=4).
\label{61aa}
\ee

\subsection{More than two `magnetic' fields}

The case $q\ge 3$ is realizable only in dimensions $d\ge 6$. 
Therefore, we need, in fact, to know $J_{s,\,q}$ only for ${\rm Re}\,s\ge 3$.
In this case $J_{s,\,q}(z)$ is an entire function of $s$ for ${\rm Re}\,z>(2-q)$.
This means that it also remains entire function of s in the limit $z\to 0$.  
Therefore, we have from eq. (\ref{29}) 

\be 
J_{d/2,\,q}=(d-2)a_{d/2,\,q}+q\,d(d-2)a_{d/2-2,q-2},
\ee
\be
J'_{d/2,\,q}=(d-2)a'_{d/2,\,q}+q\,d(d-2)a'_{d/2-2,q-2}
+4q(d-1)a_{d/2,\,q-2},
\ee
where the function $a_{k,\,q}$, $a_{k+1/2,\,q}$ and $a'_{k,\,q}$ are some {\it real numerical} constants listed in the Appendix.


\subsection{The total number of negative and zero modes}

Let us sum up the results of this section. 

\paragraph{q=1.}

In the case $q=1$ there are negative modes of the operator $\Delta$. 
Therefore, the effective action becomes complex. 
The imaginary part of the effective action `counts' the negative modes in the sense that

\be
{\rm Im}\,\Gamma_{(1)}=-{1\over 2}\pi N_{\rm YM}^{(-)},
\label{53}
\ee

\ni 
where $N_{\rm YM}^{(-)}$ is the total `number' of negative modes. Comparing (\ref{52}) and (\ref{53}) we find the number of negative modes to be
\be
N_{\rm YM}^{(-)}
=\int dx \sum_{{ \alpha}>0}
(4\pi)^{-d/2}{4\over \Gamma(d/2)}H^{d/2},
\qquad (q=1).
\label{52b}
\ee
This causes, of course, the infra-red instability of the vacuum under small (quantum) fluctuations in the directions of the negative modes.

\paragraph{q=2, d=4.}

 In the case $q=2$, $d=4$ there are no negative modes but there are {\it zero modes}. The effective action is {\it real} but it has a logarithmic {\it infra-red divergency} in the limit $M\to 0$

\be
\Gamma_{(1)}(M)\stackrel{M\to 0}{=}
{1\over 2}N_{\rm YM}^{(0)}\log\,{M^2\over H}+O(1),
\label{64aa}
\ee
where $N_{\rm YM}^{(0)}$ is the `number' of zero modes. 
Comparing eqs. (\ref{61aa}) and (\ref{64aa}) we find the total number of zero modes

\be
N_{\rm YM}^{(0)}
=\int dx \sum_{{ \alpha}>0}
(4\pi)^{-2}16 H^{2},
\qquad (q=2,\ d=4)
\label{54}
\ee

The appearance of the logarithmic infra-red divergent term does not lead to  serious physical problems. Whereas the zeta-function regularizes well the ultraviolet 
divergences, one has to renormalize additionally at the very end the coupling constants of the classical action in the infra-red region.
This means that there appears an additional {\it infra-red renormalization} parameter $\lambda$.
In other words, after we have calculated $\zeta'_{\rm YM}(0,M)$ there are no ultraviolet divergences any longer, but there are logarithmic infra-red ones. Thus one has to renormalize them as well.

\paragraph{q=2, d$>$4 and q$\ge$3, d$\ge$ 6.}

In the cases $q=2, d>4$ and $q\ge 3, d\ge 6$ there are no negative and zero modes. The effective action is real and the vacuum is stable at the quantum level, i.e., under small quantum disturbances.


\section{Analysis of the effective potential}

We define the total effective potential $V(H)$ by 
\be
\Gamma=S+\Gamma_{(1)}
=\int dx\sum_{\alpha>0}V(H).
\ee

Using the results of previous section with account of 
the classical Yang-Mills action calculated on the chromomagnetic background with equal magnetic fields 

\be
S=\int dx\sum_{\alpha>0}{2q\over g^2}H^2,
\label{61}
\ee

\ni
 we find the general form of the effective potential to be
\be
V(H)={2q\over g^2}H^2+H^{d/2}\left(a\,\log\,{H\over \mu^2}
+b+c\,\log\,{H\over M^2}\right),
\ee
where $a$, $b$ and $c$ are numerical constants depending only on the dimension of the spacetime $d$ and the number of `magnetic fields' $q$.

The constants $a$, $b$ and $c$ are calculated in previous section. The form of the effective potential depends on the signs of these constants. The constant `$a$' is not equal to zero only for even $d/2\ge 2$, i.e., in dimensions $d=4 \ ({\rm mod}\, 4)$. The constant `$b$' is always different from zero and is complex in the case $q=1$. The constant $c$ is not equal zero only in the case $q=2, d=4$.

Since in the case $q=1$ the effective potential becomes complex we do not study it in detail but concentrate our attention on the first nontrivial case $q=2$ that can be investigated analytically to the end.


\subsection{q=2, d=4}

In the case $q=2, d=4$ we have from previous section
\be
a={22\over 3(4\pi)^{2}},
\ee
\be
c=-{8\over (4\pi)^{2}},
\ee
\be
b={2\over 3(4\pi)^{2}}\left[{\zeta'(2)\over \zeta(2)}-{\bf C}
+\log 2-\log (2\pi)+1\right].
\ee

Now one should absorb the infra-red divergency by renormalizing the coupling $g$. We define first
\be
M=\lambda \varepsilon,
\ee
where $\lambda$ is arbitrary finite dimensionful parameter and $\varepsilon\to 0$. Since $V(H)$ cannot depend on the arbitrary parameters $\mu$ and $\lambda$ one assumes usually that the coupling constant depends on them in such a way that effectively the effective potential does not depend neither on $\mu$, $\lambda$ nor on $\varepsilon$, viz. 

\be
V(H)={4\over \alpha(\Lambda)}H^2 +\beta H^2 \log\,{H\over \Lambda},
\ee
where
\be
{1\over \alpha(\Lambda)}={1\over \bar g^2(\mu, \lambda)}+{c\over 4}\log{\Lambda\over \lambda^2}
+{a\over 4}\log{\Lambda\over \mu^2}
\ee
\be
{1\over \bar g^2(\mu, \lambda)}={1\over g^2(\mu, \lambda)}-{c\over 2}\log{\varepsilon}
+b,
\ee
\be
\beta=a+c=-{2\over 3(4\pi)^2},
\ee
and $\Lambda$ is a constant scale of the `magnetic field'.

As a consequence we have {\it two} different renormalization group equations for $\bar g$ with respect to $\mu$ and $\lambda$
\be
\mu{\partial\over\partial\mu}\bar g^2=\beta_{\rm UV}(\bar g),
\label{75a}
\ee
\be
\lambda{\partial\over\partial \lambda}\bar g^2=\beta_{\rm IR}(\bar g),
\ee
where 
\be
\beta_{\rm UV}(\bar g)=-{a\over 2}\bar g^4=-{11\over 3(4\pi)^2}\bar g^4,
\ee
\be
\beta_{\rm IR}(\bar g)=-{c\over 2}\bar g^4={4\over (4\pi)^2}\bar g^4,
\ee
and an equation for $\alpha(\Lambda)=\bar g^2(\Lambda, \Lambda)$
\be
\Lambda{\partial\over\partial\Lambda}\alpha
=\left.\left(\mu{\partial\over\partial\mu}
+\lambda{\partial\over\partial \lambda}\right)\bar g^2\right|_{\mu=\lambda=\Lambda}
=\beta_{\rm tot}(\alpha),
\ee
\be
\beta_{\rm tot}(\alpha)=\beta_{\rm UV}+\beta_{\rm IR}=-{1\over 2}\beta\alpha^2={1\over 3(4\pi)^2}\alpha^2
\ee

Thus we find important properties of the beta-functions

\be
\beta_{\rm UV}<0, \qquad \beta_{\rm IR}>0.
\ee

We see that according to the usual theory the effective coupling $\bar g(\mu, \lambda)$ is asymptotically free in the high-energy limit $\mu\to\infty$, $\lambda$ being fixed. Moreover, it is asymptotically free in the limit $\lambda\to 0$, $\mu$ fixed.

One should note that the behavior of the effective constant $\alpha(\Lambda)$ for $\Lambda\to\infty$ is {\it different} from the usual one. The contribution of the infra-red divergences (zero modes) {\it changes the sign} of the total beta function 
\be
\beta_{\rm tot}>0.
\ee 
This means that the effective coupling $\alpha(\Lambda)$ is {\it not asymptotically free}!
This effect occurs whenever the constant $c$ is different from zero, i.e., {\it only} in the case $q=2, d=4$.
In the standard case $q=1, d=4$ studied in the literature $\beta_{\rm IR}=c=0$ (there is no logarithmic infra-red divergency). Therefore, one has only one renormalization parameter $\mu$ and only one standard renormalization group equation (\ref{75a}) with 
$\beta_{\rm UV}<0$. This leads then automatically to the asymptotic freedom.

The behavior of the effective potential is effected by the zero modes in a similar way. 
In our case, since $\beta<0$, it is immediately seen (see Fig. 1) that although the vacuum $H=0$ is perturbatively stable
the effective potential is unbounded from below and, hence, the perturbative vacuum
is, in fact, {\it metastable}. 

One should note again that without the contribution of the zero modes one would have $\beta=a>0$ and, hence, the perturbative vacuum would be absolutely stable (see Fig. 2). These are zero modes that break down the stability of the vacuum at the nonperturbative level --- the perturbative vacuum decays  because of the zero modes.

\begin{figure}[h]
\begin{center}
\begin{picture}(220,170)
\put(0,0){\vector(0,1){150}}
\put(-14,140){$V$}
\put(0,80){\vector(1,0){200}}
\put(190,60){$H$}
\bezier{400}(0,80)(20,81)(50,100)
\bezier{400}(50,100)(88,130)(100,130)
\bezier{400}(100,130)(145,130)(180,0)
\end{picture}
\end{center}
\caption{q=2, \ d=4, 5, 6, 7 \ ({\rm mod}\ 8)}
\end{figure}
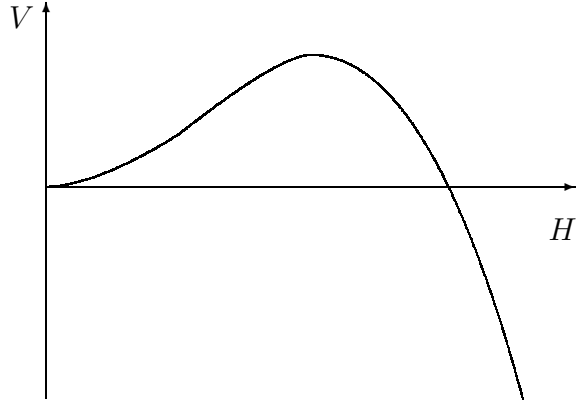


\subsection{q=2, d$>$4.}

\subsubsection{d=2k+1, 4k+2}

Consider now the case of odd dimension and odd  $d/2\ge 3$, i.e., $d=5,7,9,\dots$ and $d=6,10,14,\dots$.
Using the results of previous section we find that in both cases the constants `$a$' and `$c$' vanish

\be
a=c=0,
\ee
and the constant `$b$' can be written in the form
\be
b={2^{d/2-2}\over \sin(\pi d/4)}
(d-2)^2 {2\pi\over (8\pi^2)^{d/2}}\zeta(d/2).
\label{80aa}
\ee

\paragraph{d=8k+1, 8k+2, 8k+3.}

We see from (\ref{80aa}) that for $d=9,10,11$ (mod~8), the coefficient $b$ is positive and, therefore, the effective potential is positive and monotone (see Fig. 2). The vacuum $H=0$ gives the absolute minimum of the effective potential and is stable.

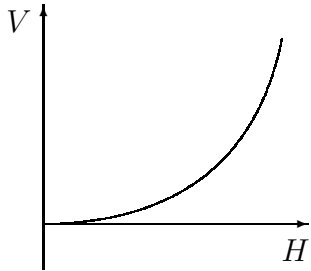
\begin{figure}[h]
\begin{center}
\begin{picture}(120,110)
\put(0,0){\vector(0,1){100}}
\put(-14,90){$V$}
\put(0,17){\vector(1,0){100}}
\put(90,3){$H$}
\bezier{400}(0,17)(76,18)(90,87)
\end{picture}
\end{center}
\caption{q=2, \ d=9, 10, 11\ ({\rm mod}\ 8)}
\end{figure}

\paragraph{d=8k+5, 8k+6, 8k+7.}

For $d=5,6,7$ (mod~8), the constant $b$ is negative. 
Qualitatively this case is similar to the case $d=4,q=2$. (See Fig. 1).
Again the effective potential is unbounded from below and the perturbative vacuum $H=0$ is {\it metastable}. 

\subsubsection{d=4k}

Finally let us consider the case of even $d/2\ge 4$, i.e., $d=8 \ ({\rm mod}\, 4)$. 
The constant `$c$'  vanish 
\be
c=0
\ee
and  the constants `$a$' and `$b$' read
\be
a=(-1)^{d/4}{2^{d/2-1}(d-2)^2\over (4\pi)^{2}\Gamma(d/2+1)}|B_{d/2}|,
\ee
\be
b=a\left({\zeta'(d/2)\over\zeta(d/2)}+{2\over d-2}-{\bf C}
+\log 2-\log(2\pi)\right).
\ee

\paragraph{d=8k+4.}

We see therefrom that
for $d=8k+4$, i.e., $d=12\ ({\rm mod}\, 8)$, the constant $a$ is negative. In this case the behavior of the effective potential repeats qualitatively that of the case $d=4$. (Fig. 1)

\subsection{q=2, d=8k.}

The only case left is that of $d=8\ ({\rm mod}\, 8)$.
This is the most interesting one. In all other cases there are no stable nontrivial, $H\ne 0$, vacuum states. Now the constant $a$ is {\it positive} and this changes radically the behavior of the effective potential.

First, it is bounded from below, i.e., there are {\it some} stable vacuums, at least the perturbative one, $H=0$. It is not difficult to see that for some values of the parameters there is indeed some nontrivial local minimum that can also provide the {\it absolute} minimum of the effective potential. The form of the effective potential is governed by the coupling constant. (See Fig. 3). One can study the behavior of the effective potential in detail and find the following.

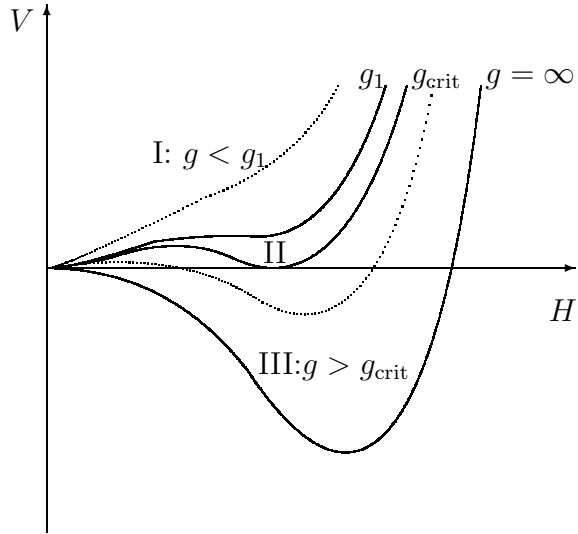
\begin{figure}[h]
\begin{center}
\begin{picture}(200,200)
\put(0,0){\vector(0,1){200}}
\put(-14,190){$V$}
\put(0,100){\vector(1,0){200}}
\put(190,80){$H$}
\put(40,140){I: $g<g_1$}
\bezier{60}(0,100)(6,100)(60,127)
\bezier{40}(60,127)(94,139)(110,169)
\bezier{500}(0,100)(10,100)(40,110)
\bezier{500}(40,110)(59,113)(80,112)
\bezier{500}(80,112)(110,112)(128,169)
\put(118,170){$g_1$}
\bezier{500}(0,100)(10,101)(24,104)
\bezier{500}(24,104)(50,113)(70,104)
\bezier{500}(70,104)(112,84)(136,169)
\put(138,170){$g_{\rm crit}$}
\put(82,102){II}
\bezier{60}(0,100)(50,108)(80,88)
\bezier{60}(80,88)(126,61)(146,169)
%
\bezier{500}(0,100)(46,100)(76,61)
\bezier{800}(76,61)(140,-35)(164,169)
\put(80,60){III:$g>g_{\rm crit}$}
\put(166,170){$g=\infty$}
\end{picture}
\end{center}
\caption{q=2, \ d=8 \ ({\rm mod}\ 8)}
\end{figure}

There are three essentially different regions of the coupling constant

\vbox{
\begin{eqnarray}
{\rm I}&:&\qquad 0<g^2<g^2_1
\nonumber\\
{\rm II}&:&\qquad g^2_1<g^2<g^2_{\rm crit}
\\
{\rm III}&:&\qquad g^2>g^2_{\rm crit}
\nonumber
\end{eqnarray}
}

\ni
where
\be
g^2_1={8(d-4)\over d a}\exp\left[{2\over d}(d-2)\right]H_0^{-(d-4)/2},
\ee

\be
g^2_{\rm crit}=2(d-4){e\over a}H^{-(d-4)/2},
\ee
where
\be
H_0=\mu^2e^{-{b\over a}}.
\ee

In the region I the effective potential is a monotone increasing function for $H>0$ and has only one minimum at $H=0$ --- the perturbative vacuum.

In region II there appear additional {\it local} minimum at some 
\be
H_{\rm min}=H_0 h_{\rm min},
\ee
where $h_{\rm min}$ is the solution of the equation
\be
h^{(d-4)/2}\left(\log h+{2\over d}\right)+{4\over d\gamma}=0,
\ee
with
\be
\gamma={a\over 4}g^2H_0^{(d-4)/2}.
\ee

\ni
The values of $H_{\rm min}$ lie in the region

\be
H_1<H_{\rm min}<H_{\rm crit},
\ee
where

\be
H_1=H_0\exp\left[-{4(d-2)\over d(d-4)}\right],
\ee

\be
H_{\rm crit}=H_0\exp\left(-{2\over d-4}\right).
\ee

This local minimum indicates that there is an additional metastable vacuum state with the energy $V_{\rm min}$ larger than that of the perturbative level, $V_{\rm min}>0$. Such metastable state must 
decay in the usual perturbative vacuum state $H=0$.

At the critical value of the coupling constant $g=g_{\rm crit}$ the effective potential has the minimum at $H_{\rm min}=H_{\rm crit}$ and the energy of this additional state becomes equal to that of the perturbative vacuum,  $V_{\rm min}=0$. This means that both states, $H=0$ and $H=H_{\rm crit}$, are physically equivalent. The true vacuum state in this case is a mixture of  the perturbative vacuum and the nonperturbative one.
In other words, at the critical value of the coupling constant a phase transition occurs: a new stable vacuum state appears and the system decays in it.

In region III the perturbative vacuum is either metastable or absolutely unstable. The minimum of the effective potential is provided by the new nonperturbative stable vacuum state $H=H_{\rm vac}$. The values of $H_{\rm vac}$ range in

\be
H_{\rm crit}<H_{\rm vac}<H_2,
\ee
where

\be
H_2=H_0 e^{-{2\over d}}.
\ee

It is interesting to note that the values of the nontrivial vacuum fields are bounded from above by $H_2$ --- there are no vacuum states with $H_{\rm vac}>H_2$. Besides, as the coupling constant $g$ increases, the vacuum energy decreases and reaches its absolutely minimal value 
\be
\min_{0<g<\infty}V_{\rm vac}=-{2a\over d e}H_0^{d/2}.
\ee


\section{Concluding remarks}

In this paper we continued the study of the low-energy behavior of the non-Abelian gauge theories initiated in \cite{Avr-jmp95a}.
We applied the covariant algebraic method for the calculation of the heat kernel of Laplace type operators developed in \cite{Avr-plb93,Avr-jmp95b} to the case of covariantly constant background fields, (\ref{2}), with $q$ `magnetic fields' of equal amplitudes, (see eqs. (\ref{16}), (\ref{17}), (\ref{26})). The heat kernels, zeta-function and the one-loop effective action are calculated explicitly in terms of the function $b_{s,\,q}(z)$ that is studied in detail in the Appendix.  The cases $q=1$ and $q=2$ for any spacetime dimension $d$ are studied analytically to the end --- the zeta-functions and the effective action are calculated exactly.

We confirmed the old result that in the case of only one `magnetic field', $q=1$, the vacuum is {\it essentially unstable}, i.e., unstable under small (quantum) disturbances.
In other words, there are {\it negative} modes that destroy the vacuum immediately --- there are no stable vacuum states with only one magnetic field.

In the case of two `magnetic fields' of equal amplitudes the effective potential is studied in detail. We distinguish two different cases $d=4$ and $d\ge 5$. 
For $d=4$ there are {\it zero} modes of the gauge fields. These zero modes do contribute to the renormalization group beta functions and {\it radically change} the asymptotic behavior of the effective coupling constant --- it is not asymptotically free any longer.
The perturbative vacuum $H=0$ is stable under small (quantum) disturbances but is unstable under classical variations of the background field.

For $d\ge 5,6,7,12\, ({\rm mod}\, 8)$ the vacuum is again metastable (Fig. 1). For $d=9,10,11\, ({\rm mod}\, 8)$ the perturbative vacuum $H=0$ is absolutely stable (Fig. 2). For $d=8\, ({\rm mod}\, 8)$ and sufficiently large coupling constant there appears some  nonperturbative stable vacuum state $H=H_{\rm vac}\ne 0$ with negative vacuum energy $V_{\rm vac}<0$ (Fig.3).

One should note that our analysis is applicable directly to Euclidean 
field theory. In case of spacetime of Lorentzian signature there is a more strong restriction $d\ge 2q+1$ --- there must be enough space to put $q$ magnetic fields. That is why our results are not applicable to the case of Lorentzian spacetimes of dimension $d=2q$ \cite{Avr-jmp95a}, e.g. for $q=2, d=4$.


\section*{Acknowledgements}

This work was supported in part by the Alexander von Humboldt Foundation and
the Deutsche Forschungsgemeinschaft. A financial support of the Naples Section of the INFN, where a part of this work was done, is gratefully acknowledged. I would like also to thank R. Musto for many fruitful discussions. I am especially indebted to G. Esposito for arranging my visit to Naples and hospitality.


\section*{Appendix. \ The function $b_{s,\,q}(z)$}
In this appendix we study the function $b_{s,\,q}(z)$ defined by

\be
b_{s,\,q}(z)={1\over\Gamma(-s)}\int\limits_0^\infty d t\,t^{-s-1+q}{e^{-tz}\over \sinh^q t}.
\label{a1}
\ee

It is not difficult to show that for ${\rm Re}\, z>-q$, the 
integral (\ref{a1}) converges in the region ${\rm Re}\,s<0$ and defines an {\it entire} 
function of $s$ by  analytical continuation. In particular, in the region ${\rm Re}\, s<N$, 
$N$ being some natural number, it is defined by

\be
b_{s,\,q}(z)={(-1)^N\over \Gamma(-s+N)}
\int\limits_0^\infty d t\,t^{-s-1+N}
{\partial^N\over\partial t^N}
\left[e^{-tz}\left({t\over \sinh t}\right)^q\right].
\label{a30}
\ee

For non-positive integer $s\ne 0,1,2,\dots$ the function $b_{s,\,q}(z)$ as a function of $z$ has a branching point $z=-q$. It is analytic function of $z$ in the complex plane cut along the real axis from $z=-q$ to $-\infty$.

At positive integer points $s=0,1,2,\dots$,  exactly in the same manner as in \cite{Avr-npb91}, we find that the values of the entire function $b_{s,\,q}(z)$ are 
determined by the Taylor expansion of the integrand

\begin{equation}
b_{k,\,q}(z)=\left.\left(-{\partial\over\partial t}\right)^{k}
\left[e^{-tz}\left({t\over\sinh t}\right)^q\right]
\right\vert_{t=0}, \qquad (k=0,1,2,\dots)
\label{a31}
\end{equation}

For half-integer $s=k+1/2$, by choosing $N=k+1$ in (\ref{a30}) we get 

\be
b_{k+1/2,\,q}(z)={1\over\sqrt\pi}\int\limits_0^\infty {dt\over \sqrt t}
\left(-{\partial\over\partial t}\right)^{k+1}
\left[e^{-tz}\left({t\over\sinh t}\right)^q\right], \qquad (k=0,1,2,\dots)
\label{a41c}
\ee

\ni
By differentiating eq. (\ref{a30}) and choosing $N=k+1$ we also obtain
the derivative 
$$
b'_{s,\,q}(z)\equiv{\partial \over \partial s}b_{s,\,q}(z)
$$
at integer points $s=k$
\be
b'_{k,\,q}(z)=
-{\bf C} b_{k,\,q}(z)
-\int\limits_0^\infty d t\,\log\, t
\left(-{\partial\over\partial t}\right)^{k+1}
\left[e^{-tz}\left({t\over\sinh t}\right)^q\right]
\label{a43c}
\ee

For nonpositive integer $q$ the function $b_{s,\,q}(z)$ is expressed in elementary functions
\be
b_{s,\,q}(z)=2^q{\Gamma(-q+1)\Gamma(-s+q)\over\Gamma(-s)}
\sum\limits_{0\le n\le -q}{(-1)^n\over n!}{(z+q+2n)^{s-q}\over \Gamma(-q-n+1)}
\ee 
$$
(q=0,-1,-2,\dots).
$$

In particular,
\be
b_{s,\,0}(z)=z^s,
\ee
\be
b_{s,\,-1}(z)={1\over 2(s+1)}\left[(z+1)^{s+1}-(z-1)^{s+1}\right],
\ee
etc.

For $q\ge 1$ one can expand $1/\sinh^q t$ in the powers of exponents to get an infinite series representation
\be
b_{s,\,q}(z)=2^q(-1)^q{\Gamma(s+1)\over\Gamma(s-q+1)\Gamma(q)}
\sum\limits_{n\ge 0}{\Gamma(q+n)\over n!}(z+q+2n)^{s-q}
\label{aa20}
\ee
Using the function (see \cite[p. 27]{Erdelyi})
\be
\Phi(\lambda, p, v)={1\over\Gamma(p)}\int\limits_0^\infty dt\,t^{p-1}
{e^{-vt}\over 1-\lambda e^{-t}}=\sum\limits_{n\ge 0}{\lambda^n\over (n+v)^p}
\ee
one has finally
\be
b_{s,\,q}(z)=(-1)^q2^s{\Gamma(s+1)\over\Gamma(s-q+1)\Gamma(q)}
\left.\left({\partial\over\partial\lambda}\right)^{q-1}\left[\lambda^{q-1}
\Phi\left(\lambda,q-s,{z+q\over 2}\right)\right]\right|_{\lambda=1}
\ee
Some properties of the function $\Phi(\lambda, p, v)$ as well as its connection with the Riemann and Hurwitz zeta-functions are described in \cite{Erdelyi}.

Let us study now the boundary value of the function $b_{s,\,q}(z)$ and its derivatives as $z\to 0-i\varepsilon$

\be
a_{s,\,q}\stackrel{\rm def}{=}\lim_{z\to 0-i\varepsilon}b_{s,\,q}(z),
\qquad
a'_{s,\,q}\stackrel{\rm def}{=}\lim_{z\to 0-i\varepsilon}b'_{s,\,q}(z).
\ee
The values of the function $b_{s,\,q}(z)$ at positive integer  points given by (\ref{a31}) are polynomials in $z$
\be
b_{k,\,q}(z)=\sum\limits_{0\le n\le k} {k\choose n}z^{k-n}a_{n,\,q},
\ee
where
\be
a_{k,\,q}=\left.\left(-{\partial\over\partial t}\right)^{k}
\left({t\over\sinh t}\right)^q
\right\vert_{t=0}, \qquad (k=0,1,2,\dots)
\label{a31a}
\end{equation}

 Taking the limit $z\to 0$ and observing that the Taylor series of the function $(t/\sinh t)^q$ is a power series in $(-t^2)$ with {\it positive} coefficients we find 

\be
a_{2k+1,\,q}=0 \qquad (k=0,1,2,\dots)
\label{a38b}
\ee
and the constants $a_{2k,\,q}$ have the property
\be
a_{2k,\,q}=(-1)^k|a_{2k,\,q}|.
\ee

The values of function $a_{s,\,q}$ at half integer points $s=k+1/2$ and its derivative $a'_{s,\,q}$ at integer points $s=k$ depend crucially on $q$. 
For nonpositive $q$ the function $b_{s,\,q}(z)$ is not analytic at the point $z=0$. Therefore, one has to take into account the infinitesimal imaginary part in $z\to 0-i\varepsilon$.

We have for $q=-1$
\be
a_{k+1/2,\,-1}={1\over 2k+3}
\left[1+i(-1)^{k+1}\right],
\ee
\be
a'_{k,\,-1}=-{1\over 2(k+1)^2}
\left[1+(-1)^k\right]+(-1)^{k+1}i\pi {1\over 2(k+1)}.
\ee
and for $q=0$
\be
a_{k+1/2,\,0}=a'_{k+1,\,0}=0, \qquad (k=0,1,2,\dots)
\ee
For $k=0$ $a'_{k,\,0}$ is not well defined because the derivative $b'_{s,\,0}(z)$ at $s=0$ is singular
\be
b'_{0,\,0}(z)=\log z+o(z).
\ee

As we already mentioned above $b_{s,\,q}(z)$ is an entire function of $s$ for ${\rm Re}\,z>-q$. 
 This means that for $q\ge 1$ the function $b_{s,\,q}(z)$ also remains analytic in the limit 
$z\to 0$, i.e., $a_{s,\,q}$ is entire function of $s$ for $q\ge 1$.
Therefore, for $q\ge 1$ we can just put $z=0$ in eqs. (\ref{a41c}) and (\ref{a43c}) to obtain

\be
a_{k+1/2,\,q}={1\over\sqrt\pi}\int\limits_0^\infty {dt\over \sqrt t}
\left(-{\partial\over\partial t}\right)^{k+1}\left({t\over \sinh t}\right)^q
\qquad (k=0,1,2,\dots),
\label{a41b}
\ee

\ni
and

\begin{equation}
a'_{k,\,q}=
-{\bf C} a_{k,\,q}
-\int\limits_0^\infty d t\,\log\, t
\left(-{\partial\over\partial t}\right)^{k+1}\left({t\over \sinh t}\right)^q
\label{a43d}
\end{equation}

$$
(k=0,1,2,\dots).
$$
We see that $a_{k+1/2,\,q}$ and $a'_{k,\,q}$ are some {\it real} numerical constants. 

There are two important particular cases when the integral 
\be
a_{s,\,q}={1\over\Gamma(-s)}\int\limits_0^\infty d t\,t^{-s-1+q}
{1\over \sinh^q t}.
\label{aa1}
\ee
can be calculated analytically for $q\ge 1$ too, namely, $q=1$ and $q=2$.
Substituting $z=0$ in eq. (\ref{aa20}) we obtain for $q=1$
\be
a_{s,\,1}=-2s(1-2^{s-1})\zeta(-s+1)
\ee
and for $q=2$
\be
a_{s,\,2}=2^s s(s-1)\zeta(-s+1).
\ee

Using the property \cite{Erdelyi} 
\be
\zeta(1-s)=2(2\pi)^{-s}\cos\left({\pi\over 2} s\right)\Gamma(s)\zeta(s)
\ee
we obtain the values of the functions $a_{s,\,1}$ and $a_{s,\,2}$ and their derivatives at half-integer points and integer points

\be
a_{k+1/2,\,1}=(-1)^{[(k+1)/2]}2^{3/2}(2^{k-1/2}-1)
(2\pi)^{-k-1/2}\Gamma(k+3/2)\zeta(k+3/2)
\ee

\be
a'_{2k+1,\,1}=(-1)^{k+1}(2^{2k}-1)(2k+1)!(2\pi)^{-2k}\zeta(2k+1)
\ee

\bea
a'_{2k,\,1}&=&(-1)^{k}4(2^{2k-1}-1)(2k)!(2\pi)^{-2k}\zeta(2k)
\nonumber\\
&&
\times
\left[{\zeta'(2k)\over\zeta(2k)}+\Psi(2k+1)
+{2^{2k-1}\over 2^{2k-1}-1}\log 2-\log(2\pi)
\right]
\eea

\be
a_{k+1/2,\,2}=(-1)^{[(k+1)/2]}(2k-1)2^{k}
(2\pi)^{-k-1/2}\Gamma(k+3/2)\zeta(k+3/2)
\ee
\be
a'_{2k+1,\,2}=(-1)^{k+1}2^{2k}2k(2k+1)!(2\pi)^{-2k}\zeta(2k+1)
\ee
\bea
a'_{2k,\,2}&=&
(-1)^{k}2^{2k+1}(2k-1)(2k)!(2\pi)^{-2k}\zeta(2k)
\nonumber\\
&&
\times
\left[{\zeta'(2k)\over\zeta(2k)}+\Psi(2k+1)
+\log 2-\log(2\pi)+{1\over 2k-1}
\right].
\eea



\begin{thebibliography}{999}

\bibitem{Savvidy} G. K. Savvidy, Phys. Lett.  {\bf B 71}, 133 (1977).

\bibitem{Pagels} H. Pagels and E. Tomboulis, Nucl. Phys. {\bf B 143}, 485 (1978).

\bibitem{Nielsen78} N. K. Nielsen and P. Olesen, Nucl. Phys. {\bf B 144}, 376 (1978).

\bibitem{Nielsen81} N. K. Nielsen, Amer. J. Phys., {\bf 49}, 1171 (1981).

\bibitem{Olesen} P. Olesen, Phys. Scr. {\bf 23}, 1000 (1981).

\bibitem{Abbot} L. F. Abbot, Nucl. Phys. {\bf B 185}, 189 (1981).

\bibitem{Consoli} M. Consoli and G. Preparata, Phys. Lett. {\bf B 154}, 411 (1985).

\bibitem{Preparata} G. Preparata, Nuov. Cim. {\bf A96}, 366  (1986).

\bibitem{Avr-jmp95a}I. G. Avramidi, J. Math. Phys. {\bf 36} (1995) 1557

\bibitem{Avr-leipz} I. G. Avramidi, {\it Effective potential in Yang-Mills theory and the stability of the chromomagnetic vacuum}, 
Proc. IIIrd
Workshop "Quantum Field Theory under the Influence of External Conditions", 
Leipzig, 18-22 Sept. 1995, to appear

\bibitem{DeWitt} B. S. De Witt, {\it Dynamical Theory of Groups and Fields},
(Gordon and  Breach, New York, 1965).

\bibitem{Dowker} J. S. Dowker and R. Critchley, 
Phys. Rev. {\bf D13}, 3224 (1976).

\bibitem{Hawking} S. Hawking, Comm. Math. Phys. {\bf 55}, 133 (1977).

\bibitem{Elizalde} E. Elizalde, {\it Ten physical applications of spectral zeta functions}, (Springer, Berlin, 1995).

\bibitem{Avr-plb93} 
I. G. Avramidi, Phys. Lett. B {\bf 305} (1993) 27.

\bibitem{Avr-jmp95b}
I. G. Avramidi, J. Math. Phys. {\bf 36} (1995) 5055.

\bibitem{Avr-plb94} I. G. Avramidi, Phys. Lett. B {\bf 336} (1994) 171.

\bibitem{Avr-jmp96a} I. G. Avramidi, J. Math. Phys. {\bf 37} (1996) 374.

\bibitem{Avr-win} I. G. Avramidi, 
{\it  New algebraic methods for calculating 
the heat kernel and the effective action in quantum 
gravity and gauge theories},
in: {\it `Heat Kernel Techniques and Quantum Gravity'}, 
Ed. S. A. Fulling, 
{\it Discourses in Mathematics and Its Applications},  
(College Station, Texas: Department of Mathematics, Texas A\& M University,
1995), pp.~115--140.

\bibitem{Avr-qgr6} I. G. Avramidi, {\it Covariant approximation schemes 
for calculation of the heat kernel in quantum field theory}, 
University of Greifswald (September, 1995), hep-th/9509075, Proc. VI
Int. Seminar ``Quantum Gravity'', Moscow, June 12--19, 1995, to appear.

\bibitem{Avr-thess}
I. G. Avramidi, {\it Nonperturbative methods 
for calculating the heat kernel}, 
Proc. Int. Workshop `Global Analysis, Differential Geometry 
and Lie Algebras', Thessaloniki, Greece, Dec. 15-17, 1994;
hep-th/9602169;
to appear in: Algebras, Groups and Geometries (1996).

\bibitem{Avr-npb91} I. G. Avramidi, Nucl. Phys. B {\bf 355} (1991) 712.

\bibitem{Erdelyi} A. Erd\'elyi, W. Magnus, F. Oberhettinger and F. G. Tricomi,
{\it Higher Transcendental Functions}, (McGraw-Hill, New York, 1953), vol. I.

\end{thebibliography}

\end{document}